\documentclass[aps,twocolumn,groupedaddress]{revtex4-1}
\usepackage{amsmath}
\usepackage{graphicx}
\usepackage{float}

\usepackage{hyperref}
\hypersetup{
    bookmarksnumbered=true, 
    unicode=false, 
    pdfstartview={FitH}, 
    pdftitle={}, 
    pdfauthor={}, 
    pdfsubject={}, 
    pdfcreator={}, 
    pdfproducer={}, 
    pdfkeywords={}, 
    pdfnewwindow=true, 
    colorlinks=true, 
    linkcolor=blue, 
    citecolor=blue, 
    filecolor=blue, 
    urlcolor=blue 
}

\begin{document}


\title{Effects of electrode surface roughness on motional heating of trapped ions}


\author{Kuan-Yu Lin, Guang Hao Low, and Issac L. Chuang}
\affiliation{MIT-Harvard Center of Ultracold Atoms, Department of Physics, Massachusetts Institute of Technology, Cambridge, Massachusetts 02139, USA}


\date{\today}

\begin{abstract}
Electric field noise is a major source of motional heating in trapped ion quantum computation. While the influence of trap electrode geometries on electric field noise has been studied in patch potential and surface adsorbate models, only smooth surfaces are accounted for by current theory. The effects of roughness, a ubiquitous feature of surface electrodes, are poorly understood. We investigate its impact on electric field noise by deriving a rough-surface Green's function and evaluating its effects on adsorbate-surface binding energies. At cryogenic temperatures, heating rate contributions from adsorbates are predicted to exhibit an exponential sensitivity to local surface curvature, leading to either a large net enhancement or suppression over smooth surfaces. For typical experimental parameters,
orders-of-magnitude variations in total heating rates can occur depending on the spatial distribution of absorbates.
Through careful engineering of electrode surface profiles, our results suggests that heating rates can be tuned over orders of magnitudes. 
\end{abstract}

\pacs{}

\maketitle

\section{Introduction}

Laser cooled trapped ions are a well-established candidate for implementing quantum computation~\cite{qc2011wineland}. However, decoherence remains a primary obstacle to the scalability of such systems. Motional heating from  electric field noise~\cite{turchette2000heating} in particular is especially detrimental to the multi-qubit operations required for universal quantum computation. It is thus imperative that its origins are well-understood in overcoming this problem.

Significant progress has been made in the understanding the origins and factors influencing electric field noise in trapped ion systems. In experimental studies,
observed heating rate are orders of magnitude larger than predictions of Johnson noise, suggesting the existence of a non-fundamental ``anomalous heating"~\cite{turchette2000heating}. Indeed, the $d^{-4}$ scaling of heating rates, with ion-electrode distance $d$, is in general agreement with predictions of uncorrelated fluctuating surface sources~\cite{monroe2006,hite2006}.
Furthermore, the reduction of heating rates by a factor of $\sim100$ after \textit{in situ} Ar$^{+}$ bombardment~\cite{hite2006} and by a factor of $\sim2$ after pulsed laser cleaning~\cite{allcock2011} suggests that adsorbed impurities are a primary sources of surface fluctuations. Combined with 
 the measured exponential suppression of heating rates with decreasing temperature~\cite{labaziewicz2008,labaziewicz2008temp}, 
a compelling physical model for electric field noise is thus thermally activated dipole fluctuations of adsorbed atoms or molecules~\cite{safavi2011,graphenecoat2014}.

It is known that details in the fabrication process of surface electrode traps play a strong role in measured heating rates, particularly at cryogenic temperatures~\cite{labaziewicz2008,graphenecoat2014}. Celebrated works include the recognition that effects such as electrode geometry can play a strong role in the distance scaling of heating rates~\cite{lowgeo2011}, and that the scaling law of heating rates at low temperatures is of the form $\exp(-T_0/T)$~\cite{labaziewicz2008,labaziewicz2008temp}, with activation energy $T_0\sim 100K$. However, not all parameters influencing this are understood. One feature ubiquitous across all such traps and particularly poorly controlled is surface roughness, but a systemic study into its effects remains lacking. Being a geometric feature, surface roughness deserves consideration. Indeed, a rough estimate suggests that roughness could alter $T_0$ by $\sim10\%$, thus leading to dramatic changes of heating rates in the cryogenic regime. 

In this work, we theoretically model the effects of electrode surface roughness on trapped ion heating rates driven by adsorbate dipole fluctuations~\cite{safavi2011}. We solve the rough surface Green's function perturbatively and apply it to find that roughness strongly affects the adsorbate-surface interaction potential. This greatly influences the strength of fluctuations and the spatial distribution of noise sources, and hence predicted heating rates. Our focus on these effects leads to a more detailed understanding of the origins of electric field noise, improving on prior works where noise sources are assumed to be identical and uniformly distributed on a smooth surface.

We find that the heating rates are exponentially enhanced or suppressed depending on the root-mean-square surface curvature -- a measure of roughness -- and the detailed spatial distribution of adsorbates. For example, in the regime where the number density of adsorbate is large, or the adsorbate-surface system is not in thermal equilibrium, a uniform density distribution of adsorbates results and leads to a predicted enhancement of heating rates over a smooth surface. Conversely, in the case of a sparse spatial distribution of adsorbates at thermal equilibrium, a suppression of heating rates is possible. These effects are particularly prevalent at low temperatures, and are strongly influenced by the profile of surface roughness.

We review in Sec.~\ref{2} the mechanism through which electric field noise is generated by adsorbate dipole fluctuations, and define surface roughness. In Sec.~\ref{3}, the effects of surface roughness on this mechanism is evaluated systematically by obtaining the rough surface Green's function in Sec.~\ref{3a} and calculating its impact on the adsorbate-surface interaction potential in Sec.~\ref{3b}. The consequences of this modified potential are studied in Sec.~\ref{4}, with two primary effects. First in Sec.~\ref{4a}, the dipole fluctuation spectral density of adsorbates is found to be highly sensitive to local surface curvature. Second in Sec.~\ref{4b}, the spatial distribution of adsorbates is shown to correlate with the local adsorbate-surface binding energy. These effects are compounded in Sec.~\ref{4c} to obtain heating rates averaged over expected distributions of surface roughness and adsorbate distributions. Additional discussion and further work is considered in Sec.~\ref{5}.

\section{Model}\label{2}
We briefly review the well-studied model of ion trap motional heating due to electric field noise~\cite{turchette2000heating,brownnutt2014ion} in Sec.~\ref{Sec:Model-Heating}. This electric field noise is assumed to arise from adsorbate dipole fluctuations~\cite{safavi2011,brownnutt2014ion} and we highlight the dominant factors that modulate its contribution to the electric field noise spectral density. The mechanism behind these dipole fluctuations is outlined in Sec.~\ref{Sec:Model-Mechanism}, and all these factors are impacted by electrode surface roughness, defined in Sec.~\ref{Sec:Model-Roughness}.

\subsection{Dipole fluctuation induced heating}
\label{Sec:Model-Heating}
Consider a single trapped ion with charge $q$, mass $m$, and secular frequency $\omega$. A fluctuating electric field $\vec E$ at the position of the ion drives excitation from the motional ground state of the ion wavepacket to its first excited state. The rate of this transition defines the heating rate~\cite{turchette2000heating}
\begin{equation}
\label{Eq:HeatingRate}
\Gamma_{0\rightarrow 1}=\frac{q^2}{4m\hbar\omega}S_{E_k}(\omega),
\end{equation}
where $S_{E_k}$ is the corresponding electric field noise spectral density in the k-th direction. Due to this direct proportionality, we will refer to $\Gamma_{0\rightarrow1}$ and $S_{E_k}$ interchangeably in the following. This quantity
\begin{equation}
S_{E_k}(\omega)\equiv2\int_{-\infty}^{\infty}\langle E_k(t)E_k(t+\tau) \rangle_te^{i\omega\tau} d\tau=2|E_k(\omega)|^2,
\label{noisespecdensity}
\end{equation}
where $\langle\rangle_t$ represents time-averaging, is established
via the Wiener-Khinchin theorem~\cite{davenport1958random}, which relates the autocorrelation function and the power spectral density of a signal.

Dipole fluctuations are widely believed to be a dominant source of electric field noise. In this model, the generated electric field $E_k(\omega)$ at ion position $\vec{r}$ is~\cite{lowgeo2011}
\begin{equation}
E_k(\omega)=\sum_{i} \frac{\partial}{\partial \vec{n}'_i}{\partial_ {\vec{r}_k}}G({\vec{r}\,}'_i,\vec{r})\mu_i(\omega),
\label{efieldfreq}
\end{equation}
where $\vec{n}'_i$ is the unit vector normal to the surface at location ${\vec{r}\,}'_i$ of the $i$-th adsorbate, $G$ is Green's function that satisfies $\nabla^2 G({\vec{r}\,}',\vec{r})=\delta({\vec{r}\,}'-\vec{r})$ with the boundary condition $G({\vec{r}\,}',\vec{r})=0$ when ${\vec{r}\,}'$ is on the electrode surface, and $\mu_i(\omega)$ represents dipole fluctuations in the frequency domain.

To zeroth order, the interaction between adatom dipoles is neglected. This produces a completely uncorrelated dipole spectrum
\begin{equation}
2\langle \mu_i(\omega)\mu_j^*(\omega) \rangle=\delta_{ij}S_{\mu_i}(\omega),
\label{dipolecorre}
\end{equation} 
where $\langle \rangle$ is the ensemble average, and $S_\mu(\omega)$ is the power spectral density of dipole fluctuations, defined in the same way as Eq.~\ref{noisespecdensity}. Combining Eqs.~\ref{noisespecdensity},~\ref{efieldfreq},~\ref{dipolecorre}, we obtain the net electric field spectral density
\begin{equation}
S_{E_k}(\omega)=\sum_{i} S_\mu({\vec{r}\,}'_i,\omega) \left|\frac{d}{d\vec{n}'_i}{\partial _{\vec{r}_k}}G({\vec{r}\,}'_i,\vec{r})\right|^2.
\label{sumse}
\end{equation}

In typical experiments, the spacing between adatoms $\sim$10nm is much smaller than the ion-electrode spacing $10$--$100\mu$m~\cite{bruzewicz2015data}. Thus we take the continuum limit by replacing the sum in Eq.~\ref{sumse} with an integral over the electrode surface $R$:
\begin{align}\nonumber
S_{E_k}(\omega)&=\int_{{\vec{r}\,}'\in R}\sigma_{\mu}({\vec{r}\,}') S_\mu({\vec{r}\,}',\omega) \left|\frac{d}{d\vec{n}'_i}{\partial _{\vec{r}_k}}G({\vec{r}\,}'_i,\vec{r})\right|^2 d{\vec{r}\,}',\\ 
\sigma_\mu({\vec{r}\,}')&=\sum_i\delta({\vec{r}\,}'-{\vec{r}\,}'_i)
\label{hrate},
\end{align}
where $\sigma_\mu({\vec{r}\,}')$ represents the local density of adsorbates at ${\vec{r}\,}'$. Thus we see the three primary factors that influence the electric field noise spectrum in Eq.~\ref{hrate}, and hence heating rates: (1) the spatial distribution of adsorbates $\sigma_{\mu}$, (2) the dipole noise emission strength $S_{\mu}$, and (3) the Green's function $G$. These factors all depend on electrode roughness, which we will demonstrate in Sec.~\ref{3} and ~\ref{4}.

\subsection{Electrode-adsorbate interactions}
\label{Sec:Model-Mechanism}
The dipole spectral density depends strongly on the species of absorbate in question -- these range from organic hydrocarbon chains to single atoms~\cite{cyr1996functional}. We shall only consider better-understood physical adsorption of atoms~\cite{safavi2011,brownnutt2014ion}, or adatoms, which results from a balance between the attractive van der Waals force and the repulsive atom-wall electron exchange interaction force~\cite{hill1952}. 

The van der Waals atom-wall potential arises from the interaction of an atomic dipole with its image charge. Hence, it scales as 
\begin{align}
V(z)=-\frac{C_3}{z^3},
\end{align}
where $z$ is the atom-wall distance~\cite{bloch2005atomwallvdw}. This is balanced by the repulsive atom-wall exchange potential. Whereas the atom-atom potential is represented by the Lennard-Jones 6-12 potential~\cite{hoinkes1980atomwallpotential} with scaling $r^{-12}$, the atom-surface repulsion potential is calculated by integrating this potential over the electrode bulk in the continuum limit. For an infinite plane, one obtains the 9-3 potential~\cite{steele1974interaction}
\begin{equation}
U(z)=\frac{C_9}{z^9}-\frac{C_3}{z^3},
\label{9-3}
\end{equation}  
where $C_9$ and $C_3$ are positive parameters dependent of specific species of adatoms and electrode atoms.

This 9-3 potential holds several bound vibrational states, with the ground states localized around the minimum of the potential $U(z)$. This minimum
\begin{align}
\label{Eq:BindingEnergy}
\textbf{U} = U(z_0),
\end{align}
approximates the binding energy of the ground state, where $z_{0}$ is the classical equilibrium position of this minimum. In the following, this classical approximation is justified as we will only consider small shifts in ratios of $\textbf{U}$ with respect to local surface curvature. These states can be approximated with a local harmonic potential, which allows one to estimate the energy spacing between the ground state and the first excited state 
\begin{equation}
\nu=\left.\sqrt{\frac{1}{m}\frac{\partial^2 U}{\partial z^2}}\right|_{z_{0}}.
\label{nu}
\end{equation}
This harmonic approximation is justified so long as $\frac{\partial^2 U}{\partial z^2}$ remains relatively constant over the spatial extent of the ground state wave packet. 

At cryogenic
temperatures, adatom dynamics are well-approximated by a thermally activated two level system. The dipole fluctuation spectral density is given by a Lorentzian~\cite{dutta1981noise}:
\begin{equation}
S_\mu(\omega)=  (\langle\mu_1\rangle-\langle\mu_0\rangle)^2 \frac{2\Gamma_0}{\omega^2+\Gamma_0^2}e^{-\frac{h\nu}{kT}},
\label{emission strength}
\end{equation}
where $\langle\mu_i\rangle$ is the expectation value dipole moment for the vibrational state $|i\rangle$, $\Gamma_0$ and $\nu$ are the transition rate and frequency from the ground state to the first excited state respectively, and $T$ is electrode temperature. It has been suggested that these transitions could be induced by vibrations of electrode atoms, resulting in fluctuations of the adatom-electrode interaction potential $U(z)$~\cite{safavi2011,brownnutt2014ion,bruzewicz2015data} driving a phonon-induced transition rate
\begin{equation}
\Gamma_0 \propto\nu^4
\label{gamma}.
\end{equation}
The exact form of $\Gamma_0$ turns out to be unimportant as its variation with roughness is small compared to other effects as will be shown in Sec.~\ref{3}.

\subsection{Surface roughness}
\label{Sec:Model-Roughness}
\begin{figure}[t]
	\includegraphics[width=0.45\textwidth]{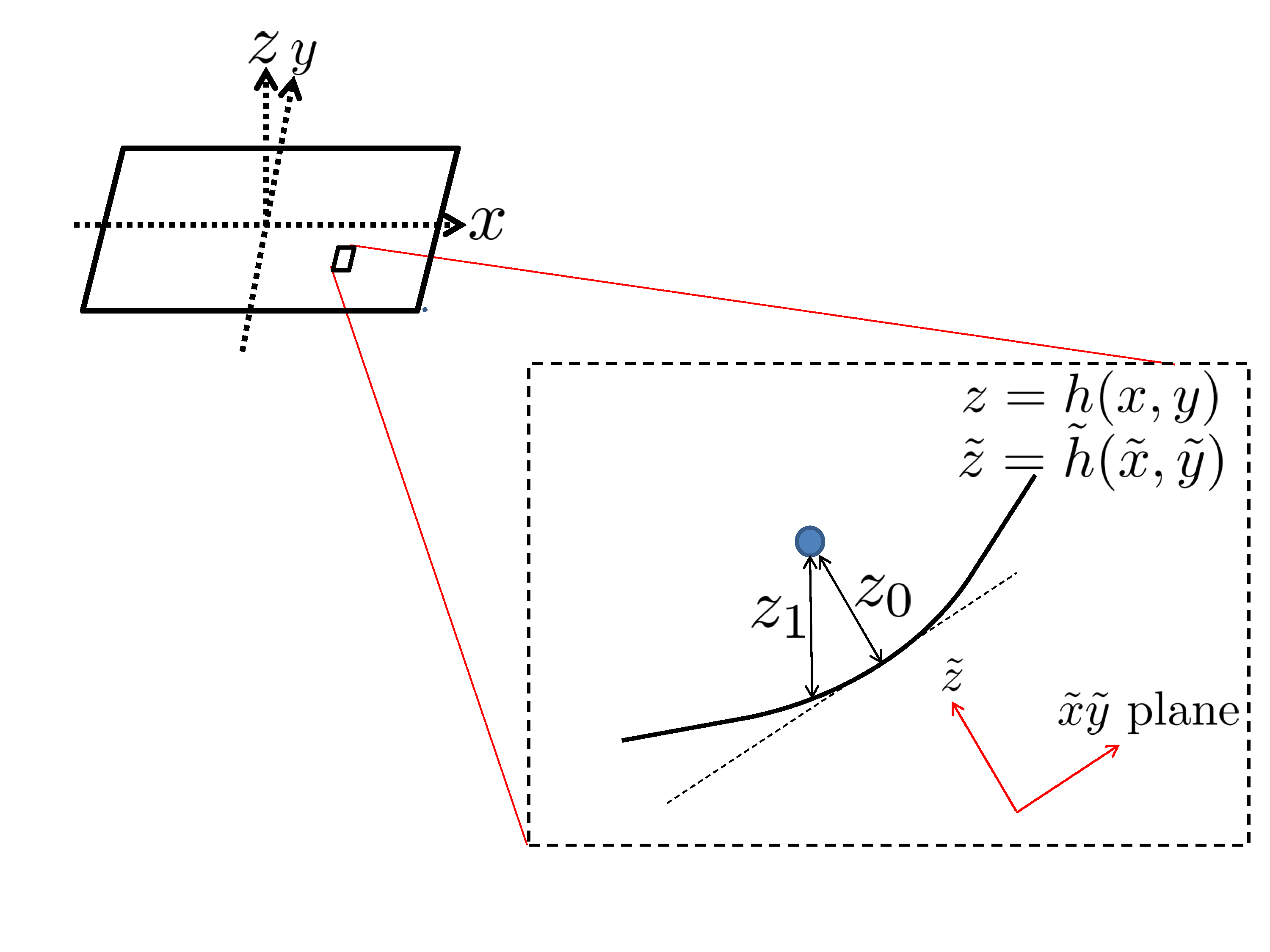}
	\caption{Cartesian coordinates defining positions on smooth and rough surfaces. The $xyz$ coordinates describe the macroscopic smooth planar geometry and is the reference against which the rough surface is defined through the height function $z=h(x,y)$. The $\tilde{x}\tilde{y}\tilde{z}$ axes describe a local coordinate system tangent to the rough surface at position $(x,y,h(x,y))$.
    }
	\label{allthesymbols}
\end{figure}

Surface electrode roughness describes height deviations from a smooth conducting surface on length scales much smaller than the gross geometry of electrode. We parameterize surface roughness and adatom positions with two Cartesian coordinate systems shown in Fig.\ref{allthesymbols}. Let us denote the hypothetical smooth surface to be the $x-y$ plane at $z=0$ where $z$ is the axis normal to the $x-y$ plane. The rough surface $R$ is thus defined through the height function $z=h(x,y)$. At any given point $(x,y)$, denote the plane tangent to the rough surface to be the $\tilde{x}-\tilde{y}$ plane, and $\tilde{z}$ the axis normal to it. 

We assume that the height function defining roughness is random in the sense of its autocorrelation function. Though this could be arbitrary, we use the very common Gaussian model in the following for concreteness
\begin{equation}
\label{Eq:RoughAutocorrelation}
\langle h(\vec{r})h(\vec{r}+\vec{v})\rangle=L^2e^{-v^2/d_0^2},
\end{equation}
where $\langle \cdot \rangle$ denotes the ensemble average over surfaces, $L$ is the root-mean-squared height of bumps on the surface, $d_0$ is the characteristic correlation length describing the width of these bumps, and $\vec{r}$, $\vec{v}$ are vectors on $z=0$ smooth plane. It is also a commonly assumed property of random surfaces that their Fourier components $h(\vec{k})=\int h(\vec{r}) e^{i\vec{k}\cdot{\vec{r}}} d\vec{r}$ are independent~\cite{patir1978numerical, longuet1957statistical}:
	\begin{equation}
	\langle h(\vec{k})h(\vec{k'})\rangle=|h(\vec{k})|^2\delta(\vec{k}+\vec{k'}).
	\label{Eq:hkcorrelation}
	\end{equation}

The surface curvature $H$ will be central to our results
\begin{equation}
H(\vec{r})=\frac{1}{2}\nabla^2 h(\vec{r}),
\end{equation}
where $\nabla^2=\partial_{x}^2+\partial_{y}^2$ is the Laplacian. In particular, we will be concerned with its probability distribution $P(H)$. For Gaussian rough surfaces, it can be proven from Eq.~\ref{Eq:RoughAutocorrelation} and \ref{Eq:hkcorrelation} and Wick's theorem that $H$ is Gaussian distributed
\begin{equation}
P(H)=\frac{1}{H_0\sqrt{2\pi}}e^{-H^2/2H_0^2}.
\end{equation}
The variance $H_0^2$ of $H$ can be computed from
\begin{equation}
H_0^2=\big\langle H(\vec{r})^2 \big\rangle=\int H(\vec{r})^2 d\vec{r}.
\label{Eq:RMS_H}
\end{equation}
By taking the Fourier transform of $H(\vec{r})$,
\begin{equation}
H(\vec{r})=\frac{1}{4\pi}\int -k^2 h(\vec{k}) e^{i\vec{k}\cdot\vec{r}} d\vec{k},
\end{equation}
applying the Wiener-Khinchin theorem,
\begin{equation}
\int H(\vec{r})^2 d\vec{r}=\frac{1}{8\pi}\int \vec{k}^4 |h(\vec{k})|^2 d\vec{k},
\label{sosA}
\end{equation}
and taking derivatives of
\begin{equation}
\int \big\langle h(\vec{r})h(\vec{r}+\vec{v})\big\rangle d\vec{r}=\frac{1}{2\pi} \int |h(k)|^2 e^{i\vec{k}\cdot\vec{v}} d\vec{k}.
\label{powerspect}
\end{equation} 
we find that
\begin{align}
4H_0^2=L^2(\nabla^2)^2e^{-|\vec{r}|^2/d_0^2}\big|_{|\vec{r}|=0}=\frac{32L^2}{d_0^4}
\end{align}
thus the RMS curvature $H_0=\frac{2^{3/2}L}{d_0^2}$. Note that though we have made assumptions on the autocorrelation function, this could in principle be directly measured.

\section{Effects of roughness on the absorbate-surface potential}\label{3}
Due to various imperfections during fabrication, roughness is a ubiquitous property of electrode surfaces and must be accounted for due to its influence on all three components in Eq.~\ref{hrate}. These are (1) the surface Green's function $\frac{d}{d\vec{n}'_i}G({\vec{r}\,}'_i,\vec{r})$ which is altered by the geometric effect of a deformed boundary; (2) the dipole emission spectrum $S_\mu({\vec{r}\,}',\omega)$ which shifts due to the change of interaction strength between adatoms and the surface; and (3) the adatom spatial density $\sigma_{\mu}({\vec{r}\,}')$ which follows the spatially varying interaction strength at thermal equilibrium. In this section, we focus on (1) and its impact on the atom-surface interaction potental. Factors (2-3) will be analyzed in Sec.~\ref{4}.

The Green's function is obtained in Sec.~\ref{3a} by solving Laplace's equation for rough surface conducting boundary conditions. This is generally difficult -- most of prior art for rough surfaces consider the scattering of electromagnetic wave in the far field limit~\cite{emscatter2009,ishimaru2000rough}. However, we require the Green's function for static sources in the near field regime. Thus, we treat the surface roughness as a small parameter in a peturbative solution with respect to the smooth surface Green's function. 

With this rough surface Green's function, we calculate in Sec.~\ref{3b} the shift in the adatom-surface interaction potential. In the presence of roughness, induced charges from the adatom are displaced to positions dependent of local topography of the surface, and therefore modify the van der Waals's interaction potential. We find that negative(positive) surface curvatures result in a weaker(stronger) van der Waals potential, which is consistent with analytical calculations for a spherical conductor/cavity in~\cite{spherevdw2011}. Furthermore, these curvatures lead to a weaker(stronger) atom-surface repulsion potential due to fewer(greater) electrode atoms contributing to atom-atom repulsion. Combining these two effects, the minima of the adatom-surface interaction potential -- the binding energy -- is correlated with the local curvature. 

\subsection{Rough Surface Green Function}
\label{3a}
Our use of rough surfaces means that traditional image charge methods are inapplicable to calculating Green's functions. Thus, we develop a perturbative solution by treating roughness as a perturbation to a smooth surface. The obtained perturbative solution allows us to calculate the change of adatom-electrode interaction potential with respect to a smooth surface, and thereafter furnishes the shift in noise spectral density of Eq.~\ref{hrate}.

We solve for the Green's functions $\mathcal{G}(\vec{r},\vec{v},\lambda)$ with the boundary condition $\mathcal{G}(\vec{r},{\vec{r}\,}',\lambda)=0$ for ${\vec{r}\,}':(x,y,z)=(r'_x,r'_y,\lambda\cdot h({\vec{r}\,}'_\perp))$ where $\lambda$ is a mathematically constructed parameter we choose with its value between 0 and 1, and ${\vec{r}\,}'_{\perp}$ is the projection of ${\vec{r}\,}'$ on the $x-y$ plane. When $\lambda=0$, the boundary condition is $\mathcal{G}({\vec{r},{\vec{r}\,}',0})=0$ on $z=0$ -- a smooth infinite plane solved by the image charge method with the well-known solution
\begin{equation}
G_0(\vec{r},\vec{v})\equiv \mathcal{G}(\vec{r},\vec{v},0)=\frac{1}{4\pi}\Big(\frac{1}{|\vec{r}-\vec{v}|}-\frac{1}{|\vec{r}-(\vec{v}-2v_z\hat{z})|}\Big),
\label{g0}
\end{equation} 
where $\vec{v}:(x,y,z)=(v_x,v_y,v_z)$ is an arbitrary point with $v_z>0$.

The known solution $G_0$ at $\lambda=0$ provides a starting point for calculating the Green's function for surface roughness $h({\vec{r}\,}')$ at $\lambda=1$. Eq.~\ref{g0} allows us to obtain the series expansion of $\mathcal{G}$:
\begin{equation}
\begin{split}
\mathcal {G}(\vec{r},\vec{v},\lambda)&=\sum_{i=0}^{\infty}\lambda^iG_i(\vec{r},\vec{v}),\\
\textrm{with }\nabla^2_{v}G_i(\vec{r},\vec{v})&=0,i\geq 1.
\end{split}
\label{seriesexp}
\end{equation}
When ${\vec{r}\,}':(x,y,z)=(r'_x,r'_y,\lambda h({\vec{r}\,}'_\perp)))$, the LHS of Eq.~\ref{seriesexp} is 0. By Taylor expanding the RHS and setting the coefficient of $\lambda^n$ to zero, we obtain equations relating higher orders $G_{n}(\vec{r},{\vec{r}\,}'_{\perp})$ with $h({\vec{r}\,}'_{\perp})$ and lower orders $G_{0}(\vec{r},\vec{v}),...,G_{n-1}(\vec{r},\vec{v})$:
\begin{equation}
\begin{split}
n=1:\; &h({\vec{r}\,}'_{\perp})\frac{\partial}{\partial {{\vec{z}\,}'}}G_0(\vec{r},{\vec{r}\,}'_{\perp})+G_1(\vec{r},{\vec{r}\,}'_{\perp})=0,\\
n=2:\;&\frac{h^2({\vec{r}\,}'_{\perp})}{2}\frac{\partial^2}{\partial {\vec{z}\,}'^2}G_0(\vec{r},{\vec{r}\,}'_{\perp})+h({\vec{r}\,}'_{\perp})\frac{\partial}{\partial {\vec{z}\,}'}G_1(\vec{r},{\vec{r}\,}'_{\perp}),\\
&+G_2(\vec{r},{\vec{r}\,}'_{\perp})=0\\
n=k:\;&\sum\limits_{i=0}^k\frac{h^{k-i}({\vec{r}\,}'_{\perp})}{(k-i)!}\frac{\partial^{k-i}}{\partial {\vec{z}\,}'^{k-i}}G_i(\vec{r},{\vec{r}\,}'_{\perp})=0.
\end{split}
\label{perturb}
\end{equation}
The $G_n(\vec{r},\vec{v})$ are solved iteratively starting from $n=1,2,...$.
To obtain $G_{n}(\vec{r},\vec{v})$ from $G_{n}(\vec{r},{\vec{r}\,}'_{\perp})$, notice that $G_{n}(\vec{r},\vec{v})\rightarrow0$ when $v_z\rightarrow\infty$ and a general solution 
\begin{equation}
G_{n}(\vec{r},\vec{v})=\frac{1}{2\pi}\int A_{n}(\vec{k'}) e^{i\vec{k'}\cdot\vec{v}_{\perp}}e^{-|\vec{k'}|v_z}  d\vec{k'},
\label{gnfull}
\end{equation} 
is obtained, where $A_n(\vec{k'})$ is defined through the boundary condition
\begin{equation}
G_{n}(\vec{r},\vec{v}_{\perp})=\frac{1}{2\pi}\int A_{n}(\vec{k'}) e^{i\vec{k'}\cdot\vec{v}_{\perp}} d\vec{k'},
\label{gfourier}
\end{equation}
where $\vec{v}_{\perp}$ is $(v_x,v_y,0)$, the projection of $\vec{v}$ onto the $z=0$ surface, and $\vec{k'}=(k_x,k_y,0)$. Observe that Eq.~\ref{gnfull} reduces to Eq.~\ref{gfourier} when $v_z=0$, so the expression in Eq.~\ref{gnfull} indeed satisfies the boundary condition. The existence of such $A_n(\vec{k'})$ arises from the invertibility of Fourier transforms, and the uniqueness of $G_n(\vec{r},\vec{v})$ is a consequence of Liouville's theorem of harmonic functions. Thus, given $G_n(\vec{r},\vec{v}_{\perp})$ and $A_n(\vec{k'})$ from Eq.~\ref{gfourier},
\begin{equation}
\frac{\partial^{m}}{\partial ({\vec{z}\,}')^{m}}G_n(\vec{r},{\vec{r}\,}'_{\perp})=\frac{1}{2\pi}\int (-|\vec{k'}|)^mA_{n}(\vec{k'}) e^{i\vec{k'}\cdot\vec{v}_{\perp}} d\vec{k'}
\label{rot},
\end{equation}
which allows the calculation of $G_{n+1}(\vec{r},\vec{v})$. After obtaining $G_n(\vec{r},\vec{v})$, $G(\vec{r},\vec{v})$ is calculated by
\begin{equation}
G(\vec{r},\vec{v})=\mathcal{G}(\vec{r},\vec{v},1)=\sum_{i=0}^{\infty}G_i(\vec{r},\vec{v}).
\end{equation}

This perturbative approach is valid so long as surface roughness is small. To be precise, we require $\frac{h^{n}({\vec{r}\,}'_{\perp})}{n!}\frac{\partial^{n}}{\partial {\vec{z}\,}'^{n}}G_i(\vec{r},{\vec{r}\,}'_{\perp})$ for any desired order $i$ to vanish for large $n$, which is satisfied if $\max{(z_0 |H|, r_z^{-1}|h|)} \lesssim 1$, where $r_z$ is the $z$-component of $\vec{r}$, and $H$ is the curvature of the rough surface.

\subsection{Change of Surface Potential to First Order}\label{3b}
We are now ready to compute the shift in the atom-surface interaction potential, which is the sum of the van der Waals attractive potential and the exchange force repulsion potential. As shown in Fig.~\ref{allthesymbols}, we place the adatom at $\vec{r}:(\tilde{x},\tilde{y},\tilde{z})=(0,0,r_{\tilde{z}})$ and approximate the local surface as parabolic -- justified in the Appendix
\begin{equation}
\tilde{h}_{p}(\tilde{x},\tilde{y})=a\tilde{x}^2+c\tilde{y}^2,
\end{equation}
where $a=\frac{1}{2}\frac{\partial^2}{\partial \tilde{x}^2}\tilde{h}$, $c=\frac{1}{2}\frac{\partial^2}{\partial \tilde{y}^2}\tilde{h}$. The axis $\tilde{x}$, $\tilde{y}$ are chosen such that $\frac{\partial^2}{\partial \tilde{x} \partial \tilde{y}}\tilde{h}=0$. The impact on potential is then calculated to first order in $a$, $c$. 

\subsubsection{Van der Waals attractive potential}
The van der Waals interaction is calculated by evaluating the interaction energy of an adatom dipole with its image charge, and then taking the expectation value of this interaction energy, assuming the adatom in its atomic ground state. The procedure is as follows: we apply the Green's function method to calculate the potential in the space above the electrode and the induced charge at the electrode surface in the case of a single charge and an electric dipole respectively; we then calculate the attraction force exerted on the dipole, which is integrated to obtain the van der Waals potential.

Consider an ion with charge $q$ placed at position $\vec{r}$ above a rough surface $R$. The potential $V(\vec{r},\vec{v},q)$ satisfies
\begin{equation}
\begin{split}
V(\vec{r},{\vec{r}\,}',q)=0 \textrm{, for } {\vec{r}\,}'\in R, \\
\nabla_{\vec{v}}^2 V(\vec{r},\vec{v},q)=\frac{q}{\epsilon_0}\delta(\vec{r}-\vec{v}),
\end{split}
\end{equation}
and therefore $V(\vec{r},\vec{v},q)= \frac{q}{\epsilon_0}G(\vec{r},\vec{v})$.

Applying Eqs.~\ref{seriesexp},\ref{perturb} by denoting $V_i(\vec{r},\vec{v},q)= \frac{q}{\epsilon_0}G_i(\vec{r},\vec{v})$, we obtain 
\begin{equation}
\begin{split}
V_0(\vec{r},\vec{v},q)&=\frac{q}{4\pi\epsilon_0}\Big(\frac{1}{|\vec{r}-\vec{v}|}-\frac{1}{|\vec{r}-(2\vec{v}_{\perp}-\vec{v})|}\Big),\\
V_1(\vec{r},\vec{v}_{\perp},q)&=-\frac{q}{2\pi\epsilon_0}\tilde{h}(\vec{v}_{\perp})\frac{r_{\tilde{z}}}{|\vec{r}-\vec{v}_{\perp}|^3},
\label{v0v1}
\end{split}
\end{equation}
where $\vec{v}_{\perp}$ is now the projection of $\vec{v}$ onto the $\tilde{z}=0$ surface. The induced charge due to a single ion $\sigma_{q}({\vec{r}\,}'),{\vec{r}\,}'\in R$ is calculated via Gauss's law:
\begin{equation}
\sigma_{q}({\vec{r}\,}')=-\epsilon_0\frac{d}{d\tilde{n'}}V(\vec{r},{\vec{r}\,}',q),
\end{equation}
where $\tilde{n}'$ is the normal vector of $R$ at ${\vec{r}\,}'$. To first order in $\tilde{h}$, $\tilde{n}'$ is approximated by $\tilde{z}$ in the subsequent calculations, which gives
\begin{equation}
\sigma_{q}({\vec{r}\,}')=-\epsilon_0\frac{d}{d\tilde{z}}V_0(\vec{r},{\vec{r}\,}'_{\perp},q)-\epsilon_0\frac{d}{d\tilde{z}}V_1(\vec{r},{\vec{r}\,}'_{\perp},q)+\mathcal{O}(h^2).
\end{equation}

The dipole-induced charge $\sigma_p({\vec{r}\,}')$ can be obtained by superposing induced charge from two opposite-signed charge at different position $\vec{r}$. For adatoms, the displacement vector $\vec{d}$, defined as the ratio between dipole and charge $\frac{\vec{p}}{q}$, has a typical value of $\sim 0.1 \AA$. Since it is much smaller than the atom-electrode distance, which is on the order of $\sim 2 \AA$, we approximate $\sigma_p$ to first order in $\vec{p}$ and $h$:
\begin{equation}
\begin{split}
\sigma_p({\vec{r}\,}')&=-\epsilon_0\frac{d}{d\tilde{n'}}(V(\vec{r}+\frac{\vec{p}}{q},{\vec{r}\,}',q)-V(\vec{r},{\vec{r}\,}',-q))\\
&=-\epsilon_0\frac{\vec{p}}{q}\cdot\nabla_{\vec{r}}\big(\frac{d}{d\tilde{z}}V_0(\vec{r_\perp},{\vec{r}\,}'_{\perp},q)+\frac{d}{d\tilde{z}}V_1(\vec{r_\perp},{\vec{r}\,}'_{\perp},q) \big).
\label{sigmap}
\end{split}
\end{equation}

The van der Waals potential is the work done moving the dipole from $\tilde{z}=\infty$ to $\vec{r}$. Thus
\begin{equation}
\textbf{V}(r_{\tilde{z}})=\int_{R=\infty}^{r_{\tilde{z}}}F_{\tilde{z}}((\tilde{x},\tilde{y},\tilde{z})=(0,0,R))dR,
\label{wd}
\end{equation}
where $F_{\tilde{{z}}}=\vec{p}\cdot\nabla (\vec{E}\cdot{\tilde{z}})$, and $\vec{E}$ is the electric field established by the induced charge on the surface:
\begin{equation}
\vec{E}(\vec{r})=\int_{r\in S} \frac{1}{4\pi\epsilon_0} \sigma({\vec{r}\,}')\frac{\vec{r}-{\vec{r}\,}'}{|\vec{r}-{\vec{r}\,}'|^3}d{\vec{r}\,}'_{\perp}+\mathcal{O}(h^2),
\label{efield}
\end{equation}
where $S$ is the parametric surface $(\tilde{x},\tilde{y},a\tilde{x}^2+c\tilde{y}^2)$. The $\mathcal{O}(h^2)$ term in Eq.~\ref{efield} comes from changing the integrating measure from ${\vec{r}\,}'$ to ${\vec{r}\,}'_{\perp}$.


Typical atomic state transition frequencies are on the order of several THz or higher. Thus in the regime where the electrode temperature is equal to or lower than room temperature, all adatoms can be assumed to be in their internal atomic ground state, and we assume the dipole fluctuations in the orthogonal directions to be independent, meaning that the expectation value of the operator 
\begin{equation}
\langle p_i p_j \rangle=d_i^2\delta_{ij}
\label{isotrop}
\end{equation}
and therefore the crossterms $p_i p_j$ in $F_{\tilde{z}}$ vanish. Combining Eqs.~\ref{v0v1},\ref{sigmap},\ref{isotrop},we obtain
\begin{equation}
\begin{split}
F_{\tilde{z}}(r_{\tilde{z}})=-\frac{1}{4\pi\epsilon_0}&\Big[\frac{3(d_{\tilde{x}}^2+d_{\tilde{y}}^2+2d_{\tilde{z}}^2)}{16{r_{\tilde{z}}}^4}\\
&+(a+c)\left(\frac{2d_{\tilde{z}}^2+3d_{\tilde{x}}^2+3d_{\tilde{y}}^2}{16{r_{\tilde{z}}}^3}\right)\Big],
\label{eq:order1f}
\end{split}
\end{equation}
which with Eq.~\ref{wd} gives the van der Waals potential
\begin{equation}
\textbf{V}(r_{\tilde{z}})=-\frac{1}{4\pi\epsilon_0}\left(\frac{d^2}{4{r_{\tilde{z}}}^3}+(a+c)\frac{d^2}{4{r_{\tilde{z}}}^2}\right),
\label{vdwfull}
\end{equation}
when $d_{\tilde{x}}^2=d_{\tilde{y}}^2=d_{\tilde{z}}^2=d^2$ is isotropic. The term $a+c$ reflects 
the mean curvature at the point $(\tilde{x},\tilde{y},\tilde{h}(\tilde{x},\tilde{y}))$. In the coordinate system $(x,y,z)$, this mean curvature is
\begin{equation}
\begin{split}
H(x,y)&=\frac{(1+h_{y}^2)h_{xx}-h_{y}h_{x}h_{xy}+(1+h_{x}^2)h_{yy}}{2(1+h_{x}^2+h_{y}^2)^{3/2}}\\
&=\frac{1}{2}(h_{xx}+h_{yy})+\mathcal{O}(h^3),
\end{split}
\label{curvature}
\end{equation}
where $h_i\equiv$ $\frac{\partial}{\partial i}h$, $h_{ij}\equiv$ $\frac{\partial^2}{\partial i\partial j}h$ with $i,j$ being the $i,j$-th directions. Thus from Eq.~\ref{vdwfull} and \ref{curvature},the van der Waal's potential to first order in $H$ in the $(x,y,z)$ coordinate system is
\begin{equation}
\textbf{V}(z)=-\frac{1}{4\pi\epsilon_0}\left(\frac{d^2}{4z^3}+H(x,y)\frac{d^2}{4z^2}\right).
\label{eq:order1vdw}
\end{equation}

\subsubsection{Adatom-surface repulsive potential}
The repulsion potential can be calculated by integrating over the bulk of electrode atoms, each of which has a repulsion potential proportional to 
$r^{-12}$ where $r$ is the distance between the electrode atom and the absorbed atom. By taking the continuum limit of electrode atoms, the repulsion potential can be calculated via the integral
\begin{equation}
\textbf{R}(r_{\tilde{z}})=\int_{\tilde{z}=-\infty}^{\tilde{h}_p(\tilde{x},\tilde{y})}\int \frac{C_r}{[(r_{\tilde{z}}-\tilde{z})^2+\tilde{x}^2+\tilde{y}^2]^6}d\tilde{x}d\tilde{y}d\tilde{z},
\label{repulint}
\end{equation}
where $C_r$ is a constant describing the strength of the $r^{-12}$ repulsion between an adatom and an electrode atom.

\begin{figure}[t]
	\includegraphics[width=0.5\textwidth]{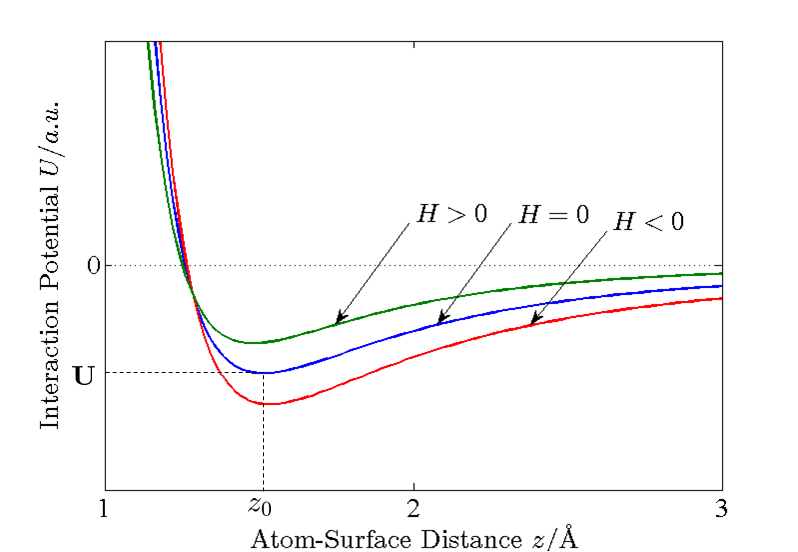}
	\caption{Qualitative plot of interaction potential $U(z)$ from Eq.~\ref{fullpot} for surfaces that are planar $H=0$, have positive curvature $H>0$, and negative curvature $H<0$. The distance scale depicted is typical for adsorbates, in this case a hydrogen adatom on a gold surface. Note in particular the direction of the shift of the binding energy, defined as the minimum $\mathbf{U}=U(z_0)$, and the transition frequency, defined through the second derivative of $U(z)$ at $z=z_0$. In the case of H-Ag interactions, $z_0\approx 1.5 \AA.$}
	\label{schpotential}
\end{figure}

To first order in the height function $\tilde{h}_{p}(\tilde{x},\tilde{y})=a\tilde{x}^2+c\tilde{y}^2$, the integral in Eq.~\ref{repulint} is approximated by
\begin{equation}
\begin{split}
\textbf{R}(r_{\tilde{z}})=\int_{\tilde{z}=-\infty}^{0}\int \frac{C_r}{[(r_{\tilde{z}}-\tilde{z})^2+\tilde{x}^2+\tilde{y}^2]^6}d\tilde{x}d\tilde{y}d\tilde{z}\\
+\int \frac{C_r \tilde{h}(\tilde{x},\tilde{y})}{({r_{\tilde{z}}}^2+\tilde{x}^2+\tilde{y}^2)^6} d\tilde{x}d\tilde{y}=\frac{\pi C_r}{45{r_{\tilde{z}}}^9}+\frac{\pi C_r(a+c)}{40{r_{\tilde{z}}}^8}.
\label{repfull}
\end{split}
\end{equation}
Using the relation $H=a+c$, we can combine Eq.~\ref{vdwfull} and Eq.~\ref{repfull} to obtain the full first order surface potential
\begin{equation}
\begin{split}
U(r_{\tilde{z}})=-\frac{d^2}{16\pi\epsilon_0}\left(\frac{1}{r_{\tilde{z}}^3}+\frac{H}{r_{\tilde{z}}^2}\right)+\frac{2\pi C_r}{90}\left(\frac{1}{r_{\tilde{z}}^9}+\frac{9H}{8r_{\tilde{z}}^8}\right)\\
=-C_3\left(\frac{1}{{r_{\tilde{z}}}^3}+\frac{H}{{r_{\tilde{z}}}^2}\right)+C_9\left(\frac{1}{{r_{\tilde{z}}}^9}+\frac{9H}{8{r_{\tilde{z}}}^8}\right).
\label{fullpot}
\end{split}
\end{equation}
Note that for a planar surface with $H=0$, the potential in Eq.~\ref{fullpot} reduces to the expected 9-3 Lennard-Jones potential. For small values of $|H| r_{\tilde{z}} \ll 1$, the sign of surface curvature produces shifts in interaction potential seen in Fig.~\ref{schpotential}. At regions of local positive curvature, the depth of potential well $\textbf{U}$ and the vibrational excitation frequency $\nu$ are larger, and vice-versa for regions of local negative curvature.

\section{Effects of roughness on heating rates}\label{4}
The dependence of the adsorbate-surface potential on local surface curvature seen in Eq.~\ref{fullpot} directly influences predicted heating rates. Specifically, the adsorbate dipole fluctuation spectral density $S_\mu$ in Eq.~\ref{emission strength} exhibits an exponential sensitivity to the transition frequency of the ground state, and we examine its dependence on roughness in Sec.~\ref{4a}. This effect is compounded by the spatial distribution of adsorbates $\sigma_\mu$ which is shown in Sec.~\ref{4a} to concentrate around regions of stronger binding energies at thermal equilibrium. We average these effects over distributions of surface roughness presented in Sec.~\ref{Sec:Model-Roughness} to obtain in Sec.~\ref{4c} the ratio of expected heating rates between rough and smooth surfaces in typical experimental regimes.


\subsection{Changes to dipole spectral density}\label{4a}
The dipole spectral density of Eq.~\ref{emission strength} is a thermally activated process and hence highly sensitive to the vibrational transition frequency $\nu$ of Eq.~\ref{nu}. For typical adatom-surface interactions, $\nu$ is on the order of $1$THz$\sim100$K. Hence in the cryogenic regime where $T\ll100$K, a small change of $\nu$ induces a large enhancement or suppression of $S_\mu(\omega)\propto \exp(-h\nu/kT)$, such as from the sign of local surface curvature $H$. To first order, 
\begin{equation}
S_\mu(\omega,H)=(\langle\mu_1\rangle-\langle\mu_0\rangle)^2\frac{2\Gamma_0}{\omega^2+\Gamma_0^2}e^{-\frac{h\nu_p(1+\mathcal{O}(H))}{kT}}.
\label{roughS}
\end{equation}
In this cryogenic regime, we treat $\mu$ and $\Gamma_0$ as constants as they only contribute linearly to the dipole spectral density, in contrast to the exponential dependence on $\nu=\nu_p(1+\mathcal{O}(H))$, where $\nu_p$ is the transition frequency for planar surface interaction.

Using the rough surface interaction potential given in Eq.~\ref{fullpot}, the dependence of the rough surface transition frequency on surface curvature can be obtained using a harmonic approximation:
\begin{equation}
\nu=\nu_p\left(1+\frac{Hz_0}{6}+\mathcal{O}(H^2)\right),
\label{roughnu}
\end{equation}
where $z_0$ is the adatom-surface equilibrium position.
From Eq.~\ref{emission strength} and~\ref{roughnu}, the ratio of $S_\mu$ between rough surface and planar surface to leading order is
\begin{equation}
\frac{S_\mu(\omega,H)}{S_\mu(\omega,0)}=\exp\left(-{\frac{h\nu_p}{kT}\frac{Hz_0}{6}}\right), 
\label{roughsmu}
\end{equation}
which, at cryogenic temperatures, shows a exponential dependence on roughness through the local surface curvature $H$.

\subsection{Changes to adatom spatial distribution}\label{4b}
\begin{figure}[t]
\includegraphics[width=0.4\textwidth]{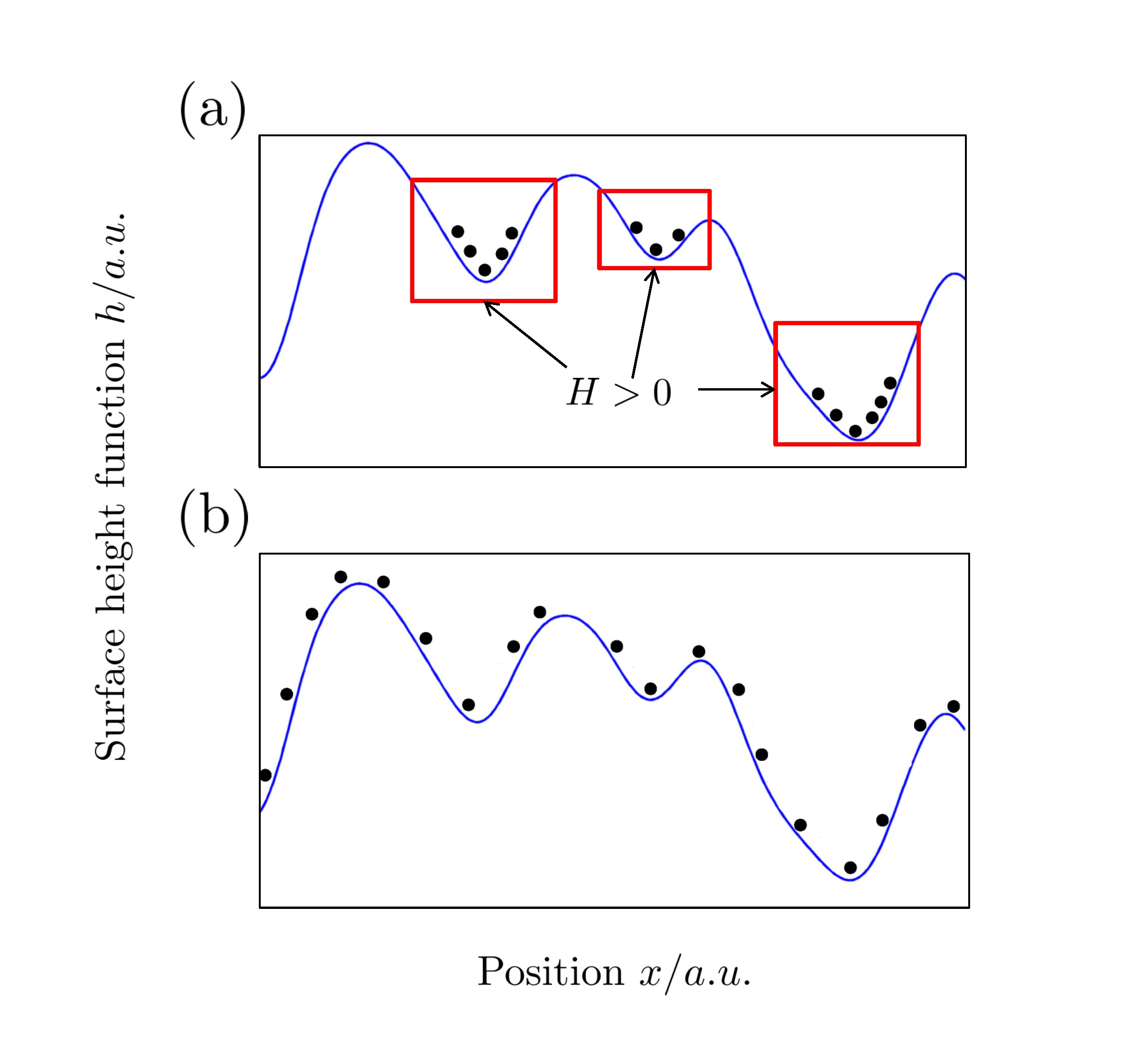}
\caption{Distribution of adatoms (dots) on a rough surface (line) in two limiting regimes. a) The thermal regime where atoms equilibrate at positions of positive curvature where the binding energy is enhanced. b) The uniform regime where atoms are uniformly distributed, such as at non-equilibrium, or if binding sites are full. The roughness (vertical axis) is exaggerated.}
\label{schdensity}
\end{figure}

We see from Eq.~\ref{roughsmu} that the dipole spectral density $S_\mu$ depends strongly on the location of  an adatom. In particular, either an exponential enhancement or suppression is possible depending on the sign of local surface curvature $H$. From Eq.~\ref{hrate}, the spatial distribution $\sigma_\mu$ of adatoms is thus critical in determining whether a net increase or decrease in heating rates over smooth surfaces is observed. For instance, suppression of the electric field noise spectral density $S_E$ occurs if all the adatoms are located at sites with positive curvature, as seen in Fig.~\ref{schdensity}(top).

The spatial distribution of adatoms is greatly affected by the binding energy $\textbf{U}$ in Eq.~\ref{Eq:BindingEnergy}. For example, adatoms at thermal equilibrium are more likely to be present at sites of higher binding energy. 
Due to the presence of roughness, this binding energy varies with location on the surface, and can induce a spatial distribution significantly different from the typically assumed uniform distribution. This dependence of $\textbf{U}$ on surface curvature can be obtained by minimizing Eq.~\ref{fullpot}:
\begin{equation}
\textbf{U}=\textbf{U}_p\left(1+\frac{15Hz_0}{16}+\mathcal{O}(H^2)\right)
\label{potdepth},
\end{equation}
where $\textbf{U}_{p}$ is the binding energy for planar surfaces. 

We consider two extreme regimes of interest for the spatial distribution of adatoms. \\
(1) The uniform regime Fig.~\ref{schdensity}(b): the spatial distribution of adatoms is approximately uniform, with constant density
\begin{align}
\sigma_\mu(\vec{r})=\sigma_\mu=\frac{N}{\int d\vec{r}},
\end{align}
where $N$ is the total number of adatoms. This arises when many adatoms are present on the surface, or a strong repulsive interaction exists between adatoms. Alternatively, the surface right after fabrication and before annealing might also be uniformly distributed, as the adatoms have not had time to reach thermal equilibrium. Given time, this uniform distribution relaxes to a Fermi-Dirac distribution through adatom diffusion~\cite{ZHDANOV1991185}, leading to the thermal regime.\\
(2) The thermal regime Fig.~\ref{schdensity}(a): we neglect the interaction between the adatoms and assume Fermi-Dirac statistics for binding sites. The local filling fraction can be written as
\begin{equation}
\theta(\vec{r})\propto \left(1+\exp{\left(\frac{-\textbf{U}(\vec{r})-\mu}{kT}\right)}\right)^{-1},
\label{fdst}
\end{equation}
where $\textbf{U}(\vec{r})$ is the local binding energy as a function of position and $\mu$ is the chemical potential. Since $\textbf{U}/k$ is typically on the order of $1000$K~\cite{xie2010,juarez2008}, the range of binding energies at cryogenic temperatures $U_0 H_{rms} z_0\gg kT$, and we assume a zero-temperature distribution of adatoms:
\begin{equation}
\theta({\vec{r}\,}')=\Theta(\mu+\textbf{U}({\vec{r}})),
\end{equation}
where $\Theta$ denotes the Heaviside step function. 
The adatom density $\sigma_\mu(\vec{r})$ is related to the filling fraction by
\begin{equation}
\sigma_\mu(\vec{r})=\frac{N\theta(\vec{r})}{\int \theta{(\vec{r})}d\vec{r}}.
\end{equation}

The behavior described by these extremes of the uniform and thermal spatial distribution of adatoms provides valuable intuition about intermediate distributions between them. Furthermore, both these extremes could occur in experiments due to the wide variation of diffusion constants for adatoms between $10^{-15}$ to $10^{-9}$m$^2$/s, which lead to timescales of $10$s to $10^7$s for a typically-sized ion trap with length dimensions $\sim0.1$mm. 

\subsection{Heating rates for random rough surfaces}\label{4c}
The heating rate is directly proportional to the spectral density of electric field noise. In the limit of small roughness, the deviation of the Green's function term in Eq.~\ref{hrate} only induces a linear dependence of roughness on heating rates, thus we focus on the dominant terms of dipole emission spectrum $S_\mu({\vec{r}\,}',\omega)$ and adatom spatial density $\sigma_{\mu}({\vec{r}\,}')$. As these have a multiplicative effect, their contribution to electric field noise can be significantly stronger than expected when two are be correlated. In order to obtain an averaged expression for heating rates, it is necessary to integrate over the surface of interest. This can be performed using a distribution $P(H)$ for the key parameter of surface curvature $H$. In the following, we apply the Gaussian rough surfaces of Sec~\ref{Sec:Model-Roughness}, where $P(H)$ is Gaussian distributed.

\begin{figure}[t]
	\includegraphics[width=0.5\textwidth]{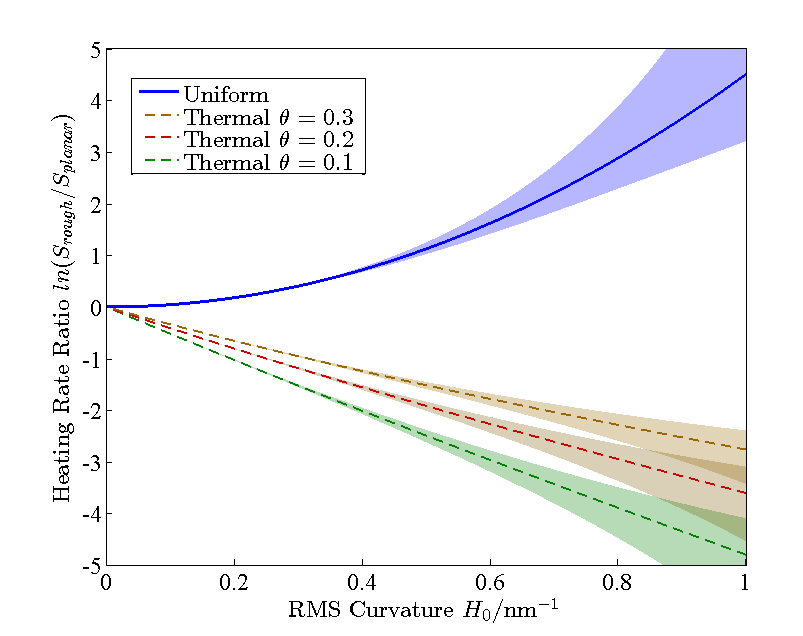}
	\caption{Ratio of Heating rate for rough surfaces $S_{\text{rough}}$ over planar surfaces $S_{\text{planar}}$ with respect to root mean squared surface curvature $H$ to first order in $H$. Normalized for the same number of adsorbates, a uniform distribution (thick) of adatoms sees a strong exponential enhancement. When the distribution relaxes to thermal equilibrium (dashed), heating rates are gradually suppressed depending on the filling fraction $\theta$ of binding sites. The shaded area depicts the estimated contribution of higher order $H^2$ terms for typical physical surfaces. Parameter values used: $h\nu_p/k_B=$200K, $z_0$=3\AA, T=4K. 
    }
	\label{heatingenhance}
\end{figure}

Given a fixed total number of adatoms on the surface, we evaluate the ratio of ensemble averaged heating rates of rough surfaces in the uniform regime $S_{\text{uniform}}$ and planar surface heating rate $S_{\text{planar}}$  is
\begin{equation}
\begin{split}
\frac{S_{\text{uniform}}}{S_{\text{planar}}}&=\int_{-\infty}^{\infty} P(H)\exp{\Big(\frac{-Hz_0}{6}\frac{h\nu_p}{kT}\Big)}dH\\
&=\exp{\left[\left(\frac{H_0z_0}{2^{3/2}\mathord{\cdot} 3}\frac{h\nu_0}{kT}\right)^2\right]}
\label{uniformratio}.
\end{split}
\end{equation}
We see a strong exponential enhancement of heating rates in Fig.~\ref{heatingenhance} which arises from adatoms at regions of negative curvature. These adatoms are more weakly bound to the surface, and consequently fluctuate exponentially more strongly -- outweighing the reduced contribution from adatoms at regions of positive curvature. Note that while we operate in the regime $|H z_0| \ll 1$, taking limits of the integration to infinity is justified as the Gaussian $P(H)$ decays exponentially more rapidly than the integrand.

The ratio between rough surface heating rates in the thermal regime $S_{\text{thermal}}$ and $S_{\text{planar}}$ once again for fixed number of adatoms is
\begin{align}
\label{denseratio}
\frac{S_{\text{thermal}}}{S_{\text{planar}}}&=\frac{1}{\theta}\int_{H(\theta)}^{\infty}  P(H)\exp\Big(\frac{-Hz_0}{6}\frac{h\nu_p}{kT}\Big)dH \\ \nonumber
&=\exp{\left[\Big(\frac{H_0z_0}{2^{3/2}\mathord{\cdot}3}\frac{h\nu_p}{kT}\Big)^2\right]}f(\theta),\\ \nonumber
f(\theta)&=\frac{1}{2\theta}\left(1-\text{erf}\left(\left(\frac{H(\theta)}{2^{1/2}H_0}+\frac{H_0z_0}{2^{3/2}\mathord{\cdot}3}\frac{h\nu_p}{kT}\right)\right)\right),
\end{align}
where $\theta=\langle \theta({\vec{r}\,}') \rangle$ is mean filling fraction, and $H(\theta)$ is such that
\begin{equation}
\int_{H(\theta)}^{\infty}P(H)dH=\frac{1}{2}\left(1-\text{erf} \left(\frac{H(\theta)}{2^{1/2}H_0}\right)\right)=\theta,
\label{Eq:Htheta}
\end{equation} 
This complicated expression simplifies at two extremes for the filling fraction.\\
(1) $\theta=1$: In this case, $H(\theta)=-\infty$, so $f(\theta)=1$ and Eq.~\ref{denseratio} is identical to that of the uniform regime. \\
(2) $\theta\ll 1$: In this case $H(\theta)$ is a large positive number, but we limit it to not too much larger than $1/z_0$ where our perturbative approach breaks down. To order $\mathcal{O}(\frac{1}{H(\theta)})$,
\begin{equation}
\frac{S_{\text{thermal}}}{S_{\text{planar}}}\approx\left({1+\frac{H_0^2z_0}{6H(\theta)}\frac{h\nu_p}{kT}}\right)^{-1}\exp\Bigg(
		-\frac{H(\theta)z_0}{6}\frac{h\nu_p}{kT}\Bigg),
		\label{Eq:smallthetaapprox}
	\end{equation}
Unlike the uniform case, a suppression of heating rates seen in Fig.~\ref{heatingenhance} occurs as all adatoms are localized to regions of positive curvature $H(\theta)>0$. From Eq.~\ref{roughnu} and Eq.~\ref{roughS}, these binding sites with deeper potential wells have larger transitional frequencies, leading to smaller dipole fluctuations.




We can also estimate the error in the ratio of heating rates arising from only considering terms linear in $H$ in this perturbative approach. This is done by obtaining an order-of-magnitude estimate for the coefficient of next-leading-order $H^2$ terms in $\textbf{U}$ and $\nu$. Through an exact calculation of the interaction between an adatom and a spherical conducting cavity in the Appendix, the ratio between second order $H^2$ and the first order $H$ terms is $CHz_0$, where $C\approx 1.19$ is a constant on the order of unity. Assuming that this magnitude of $C$ is typical for physical surfaces, we obtain the shaded region in Fig~\ref{heatingenhance} for variations in heating rates to second order with $C\in[-1.19,1.19]$.

Regardless of the exact form of the surface curvature distribution $P(H)$, a general trend is observed. Heating rates are enhanced when the adatom spatial distribution overlaps with regions of negative curvature such as in the uniform regime, and heating rates are suppressed when a large fraction of adatoms are localized to regions of positive curvature, such as in the thermal regime. Indeed, more exact results could be obtained with a more judicious choice of surface roughness autocorrelation functions. 

\section{Conclusion}
\label{5}
We have developed an analytic approach for calculating the effects of electrode surface roughness on the adbsorbate model of electric field noise, and thus the heating rates of trapped ions. Our calculations predict that, for surfaces with roughness of the scale of nanometers, an exponential suppression or enhancement of heating rates is possible, depending on the filling fraction and distribution of surface adatoms.

Our analysis provides a possible explanation for the wide spread of experimentally observed of heating rate. As roughness is poorly controlled in many experiments, possible significant factors could even include process details of the electrode trap fabrication~\cite{labaziewicz2008}. However, since the range of activation energy $\nu_0 H_0  z_0$ is on the order of 100K, we expect this roughness effect to to only be significant at cryogenic temperatures. Although we have only considered adatom adsorbates, our results motivate the investigation of other adsorbate models which could be dominant at higher temperatures. 

It would be of interest to perform a systematic study of heating rates with roughness as a control parameter. For example, surface curvatures of $H_0\sim$ 1nm$^{-1}$ have been engineered on a Ag surface\cite{gimzewski1985silver}, which from our results would correspond to a $\sim100$ fold enhancement or suppression of heating rates at cryogenic temperatures. Thus, measuring the heating rates of ions in traps with rough surfaces at temperature between $4$K and $100$K could provide for a strong experimental validation of the surface adsorbate theory of electric field noise, and would enable global probes of surface parameters through heating rate measurements.

\bibliography{main}

\appendix




\section{Justification of the parabolic approximation}
From the geometry shown in Fig.\ref{allthesymbols},
\begin{equation}
z_1=\sqrt{1+(h_x)^2+(h_y)^2}z_0=(1+\mathcal{O}(h^2))z_0,
\end{equation} 
and
\begin{equation}
\begin{split}
\frac{d}{d\tilde{n}}G_(r,r'_{\perp})&=\sqrt{1+(\tilde{h}_{\tilde{x}})^2+(\tilde{h}_{\tilde{y}})^2}\frac{d}{d\tilde{z}}G_(r,r'_{\perp})\\
&=(1+\mathcal{O}(h^2))\frac{d}{d\tilde{z}}G_(r,r'_{\perp}).
\end{split}
\end{equation}
Thus to first order, the terms $h$, $z_0$,$z_1$ and $\frac{d}{d\tilde{n}}$,$\frac{d}{d\tilde{z}}$ are interchangeable respectively. Under this assumption, Eq.~\ref{efield} can be expanded to first order in $h$ as in Eq.~\ref{Efieldgradappendix}.
\begin{widetext}
\begin{equation}
\begin{split}
\frac{\partial}{\partial \tilde{z}}E_{\tilde{z}}(\vec{r})=&\int\frac{1}{4\pi\epsilon_0} \Big\{\Big[\Big(\frac{1}{|\vec{r}-{\vec{r}\,}'_{\perp}|^3}-\frac{3r_{\tilde{z}}^2}{|\vec{r}-{\vec{r}\,}'_{\perp}|^5}\Big)
+\tilde{h}({\vec{r}\,}'_{\perp})\Big(\frac{9r_{\tilde{z}}}{|\vec{r}-{\vec{r}\,}'_{\perp}|^5}-\frac{15r_{\tilde{z}}^3}{|\vec{r}-{\vec{r}\,}'_{\perp}|^7}\Big)\Big]
\boldsymbol{\sigma_0}({\vec{r}\,}')\\
&+\Big(\frac{1}{|\vec{r}-{\vec{r}\,}'_{\perp}|^3}-\frac{3r_{\tilde{z}}^2}{|\vec{r}-{\vec{r}\,}'_{\perp}|^5}\Big)\boldsymbol{\sigma_1}({\vec{r}\,}')\Big\} d{\vec{r}\,}'_{\perp}\\
\frac{\partial}{\partial \tilde{x}}E_{\tilde{z}}(\vec{r})&=\int\frac{1}{4\pi\epsilon_0}\Big\{\Big[\frac{3(r'_{\tilde{x}}-r_{\tilde{x}})r_{\tilde{z}}}{|\vec{r}-{\vec{r}\,}'_{\perp}|^5}
+\tilde{h}({\vec{r}\,}'_{\perp})\Big(\frac{15(r'_{\tilde{x}}-r_{\tilde{x}})r_{\tilde{z}}^2}{|\vec{r}-{\vec{r}\,}'_{\perp}|^7}-\frac{3(r'_{\tilde{x}}-r_{\tilde{x}})}{|\vec{r}-{\vec{r}\,}'_{\perp}|^5}\Big)\Big]\boldsymbol{\sigma_0}({\vec{r}\,}')
+\frac{3(r'_{\tilde{x}}-r_{\tilde{x}})r_{\tilde{z}}}{|\vec{r}-{\vec{r}\,}'_{\perp}|^5}\boldsymbol{\sigma_1}({\vec{r}\,}')\Big\} d{\vec{r}\,}'_{\perp}\\
\frac{\partial}{\partial \tilde{y}}E_{\tilde{z}}(\vec{r})&=\int\frac{1}{4\pi\epsilon_0}\Big\{\Big[\frac{3(r'_{\tilde{y}}-r_{\tilde{y}})r_{\tilde{z}}}{|\vec{r}-{\vec{r}\,}'_{\perp}|^5}
+\tilde{h}({\vec{r}\,}'_{\perp})\Big(\frac{15(r'_{\tilde{y}}-r_{\tilde{y}})r_{\tilde{z}}^2}{|\vec{r}-{\vec{r}\,}'_{\perp}|^7}-\frac{3(r'_{\tilde{y}}-r_{\tilde{y}})}{|\vec{r}-{\vec{r}\,}'_{\perp}|^5}\Big)\Big]\boldsymbol{\sigma_0}(r')
+\frac{3(r'_{\tilde{y}}-r_{\tilde{y}})r_{\tilde{z}}}{|\vec{r}-{\vec{r}\,}'_{\perp}|^5}\boldsymbol{\sigma_1}(r')\Big\} d{\vec{r}\,}'_{\perp},
\end{split}   
\label{Efieldgradappendix}
\end{equation}
\end{widetext}  
where $\boldsymbol{\sigma_0}(r')$ is
\begin{equation}
\boldsymbol{\sigma_0}({\vec{r}\,}')=-\epsilon_0\frac{\partial}{\partial \tilde{z}}V_0(\vec{r},{\vec{r}\,}'_{\perp}),
\end{equation}
and $\boldsymbol{\sigma_1}(r')$ is
\begin{equation}
\boldsymbol{\sigma_1}({\vec{r}\,}')=-\epsilon_0\frac{\partial}{\partial \tilde{z}}V_1(\vec{r},{\vec{r}\,}'_{\perp})
\end{equation}
with 
\begin{equation}
V_1(\vec{r},{\vec{r}\,}'_{\perp})=-\tilde{h}({\vec{r}\,}'_{\perp})\frac{\partial}{\partial \tilde{z}}V_0(\vec{r},{\vec{r}\,}'_{\perp}).
\end{equation}

A local parabolic approximation is justified by showing consistency with randomly generated rough surfaces. We consider the form
\begin{equation}
 \begin{split}
 h_g(x,y)=\sum_{1 \leq N,M \leq 200} \big(a_{N,M}\cos(k(Nx+My))-a_{N,M}\\
 +b_{N,M}\sin(k(Nx+My)),
 \end{split}
 \end{equation}
 in which the sinusoid terms represents the Fourier-transformed coefficients of $h_g$ whose magnitude are determined by the height autocorrelation function, and the spatially constant $a_{n,m}$ terms are introduced to set $h(0,0)=0$ for convenience. $k$ is the grid size in the Fourier space for us to replace the integral by summation, in this section set to be $\frac{1}{50}\frac{1}{d_0}$. 
 
 For surfaces with a Gaussian autocorrelation function, 
 \begin{equation}
 \langle a_{N,M}^2 \rangle=\langle b_{N,M}^2 \rangle\propto e^{-\frac{N^2+M^2}{l^2}}.
 \end{equation}

We set $l=50$ in our simulation, so that the autocorrelation function takes the form 
 \begin{equation}
 \langle h_g(\vec{r})h_g(\vec{r}+\vec{v}) \rangle=L^2e^{-\frac{v^2}{d_0^2}}.
 \end{equation}
 In this case, the parabolic surface takes the form
 \begin{equation}
 h(\tilde{x},\tilde{y})=\sum_{1 \leq N,M \leq 200}-k^2(\frac{N^2}{2}\tilde{x}^2+NM\tilde{x}\tilde{y}+\frac{M^2}{2}\tilde{y}^2).
 \end{equation}
 
 Denote the perfect plane surface potential as $U_p(z)$, the parabolic surface potential as $U(z)$, and the gaussian generated surface potential as $U_g(z)$. We denote the error ratio $\epsilon$ as
 \begin{equation}
 \epsilon=\left|\frac{U(z_0)-U_g(z_0)}{U(z_0)-U_p(z_0)}\right|.
 \end{equation}

 The total error consists of the error from the van der Waal's potential and the repulsion potential. If we use a similar definition of the error ratio from the van der Waal's potential and repulsion potential,
 \begin{equation}
 \begin{split}
 \epsilon_V&=\left|\frac{\textbf{V}(z_0)-\textbf{V}_g(z_0)}{\textbf{V}(z)-\textbf{V}_p(z)}\right|\\
 \epsilon_R&=\left|\frac{R(z_0)-R_g(z_0)}{R(z_0)-R_p(z_0)}\right|,
 \end{split}
 \end{equation}
 in which, as in the case of total potential $U$, the potentials without scripts correspond to the parabolic surface $z=h(x,y)$, the ones with subscripts $g$ correspond to the gaussian generated surface $z=h_g(x,y)$, and the ones with subscript $p$ correspond to the perfect plane surface.
 
 From eq.\ref{fullpot}, 
 \begin{equation}
 \left|\frac{\textbf{V}(z)-\textbf{V}_0(z)}{R(z)-R_0(z)}\right|=\frac{8}{3}+\mathcal{O}(Az_0)
 \end{equation}
 which yields
 \begin{equation}
 \epsilon<4\max\{\epsilon_V,\epsilon_R\}
 \end{equation}
 
\begin{figure}[H]
	\includegraphics[width=0.5\textwidth]{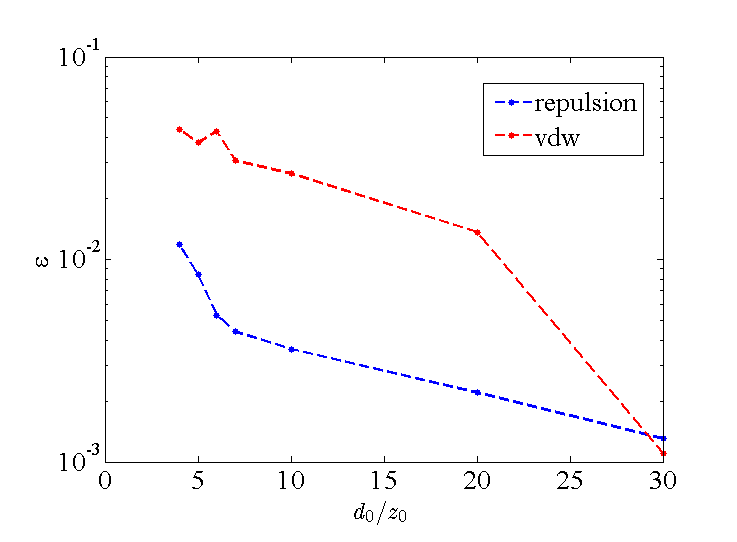}
	\caption{$\epsilon$ value for the van der Waal's potential(red) and the repulsion(blue) potential. For typical electrode surfaces, the error on the potential curve from approximating the whole surface as a parabolic surface is less than 5$\%$, and therefore the parabolic approximation is valid.}
	\label{err}
\end{figure} 
 The dependence of $\epsilon_V$ and $\epsilon_R$ on $d_0$ is plotted in Fig.\ref{err}.
 For typical metal surfaces, $d_0/z_0>20$. Therefore, in our regime of interest, $\epsilon<0.05$ and the parabolic approximation is valid in this regime.

\subsection{Estimation of the $H^2$ term in $\textbf{U}$ and $\nu$ using a spherical geometry}
Identical to the treatment in Sec.~\ref{3b}, we calculate the Van der Waal's interaction potential between a dipole and a conducting sphere through the work done in moving the dipole from infinity. 

Let the sphere be described by $x^2+y^2+z^2=R^2$, and the dipole $\vec{d}_1=(d_x,d_y,d_z)$ be located at $(0,0,R-z)$. Using the standard image charge method, one obtain at position $(0,0,\frac{R^2}{R-z})$ an image dipole $\vec{d}_2=(-\frac{R^3}{(R-z)^3}d_x,-\frac{R^3}{(R-z)^3}d_y,\frac{R^3}{(R-z)^3}d_z)$ and an image charge $\frac{d_zR}{(R-z)^2}$.

The force between dipoles $\vec{d}_1$,$\vec{d}_2$ is given by
\begin{equation}
\vec{F}_{dd}\cdot \hat{r}=\frac{3}{4\pi\epsilon_0 r^4}\left(\vec{d}_1\cdot \vec{d}_2-3(\vec{d}_1\cdot \hat{r})(\vec{d}_2\cdot \hat{r}) \right),
\label{eq:fdd}
\end{equation}
where $\vec{r}$ is the relative position $\vec{r}_1-\vec{r}_2$, and the force between a dipole $\vec{d}$ and a single charge $q$ is
\begin{equation}
\vec{F}_{dq}\cdot \hat{r}=\frac{-2q \vec{d} \cdot \hat{r} }{4\pi \epsilon_0 r^3}.
\label{eq:fdq}
\end{equation}
Using Eq.~\ref{eq:fdd} and \ref{eq:fdq}, the attraction force in the $z$ direction between the dipole and a conducting sphere is 
\begin{equation}
F_z(z)=-\frac{3(d_x^2+d_y^2+2d_z^2)}{4\pi\epsilon_0}\frac{R^3(R-z)}{z^4(2R-z)^4}+\frac{d_z^2}{4\pi\epsilon_0}\frac{2R(R-z)}{z^3(2R-z_0)^3},
\end{equation}
and thus the van der Waal's interaction to $\mathcal{O}(\left(z/R\right)^2)$ for an isotropic atom $(d_x=d_y=d_z=d)$ can be written as
\begin{equation}
\begin{split}
F_{z}(z)&=-\frac{1}{4\pi\epsilon_0}\Big[\frac{3(d_x^2+d_{y}^2+2d_{z}^2)}{16z^4}\\
&+\frac{1}{R}\left(\frac{2d_{z}^2+3d_{x}^2+3d_{y}^2}{16{r_{\tilde{z}}}^3}\right)+\frac{1}{2R^2}\left(\frac{2d_{z}^2+3d_{x}^2+3d_{y}^2}{16{r_{\tilde{z}}}^2}\right)\Big].
\end{split}
\end{equation}
By imposing the isotropic condition $d_x=d_y=d_z=d$ on the atomic state, the van der Waal's potential beceomes
\begin{equation}
\textbf{V}(z)=-\frac{1}{4\pi\epsilon_0}\left(\frac{d^2}{4z^3}+\frac{1}{R}\frac{d^2}{4z^2}+\frac{1}{R^2}\frac{d^2}{4z}\right).
\label{eq:order2vdw}
\end{equation}
The curvature of the sphere is $a+c=H=\frac{1}{R}$. Therefore, Eq.~\ref{eq:order2vdw} agrees our perturbative result in Eq.~\ref{eq:order1vdw} to first order.

The repulsion potential is the integration of $1/r^{12}$ over the bulk, which in the case of a spherical conductor is
\begin{equation}
\begin{split}
\textbf{R}(z)&=\int_{r=R}^\infty r^2dr\int d\Omega \frac{C_r}{\left(r^2+(R-z)^2-2(R-z)r\cos \theta \right)^6}\\
&=\frac{\pi}{5(R-z)}\left(\frac{z+8R}{72z^9}-\frac{10R-z}{72(2R-z)^9}\right),
\end{split}
\end{equation}
which, to $\mathcal{O}\left((z/R)^2\right)$ can be written as
\begin{equation}
\textbf{R}(z)=\frac{\pi C_r}{45}\left(\frac{1}{z^9}+\frac{9}{8Rz^8}+\frac{9}{8R^2z^7}\right).
\label{order2repul}
\end{equation}
Again, Eq.~\ref{order2repul} agrees Eq.~\ref{repfull} ti first order.

Using the harmonic approximation at the equilibrium position, we obtain the transition frequency to second order in $z/R$
\begin{equation}
\nu=\nu_p\left(1+\frac{1}{3}\frac{z}{R}+\frac{257}{648}\frac{z^2}{R^2}\right).
\end{equation}
Thus the contribution from the quadratic term is factor $C\frac{z}{R}=CHz$ larger than the first order term where $C$ is a constant on the order of unity ($\sim$1.19).
\end{document}